\documentclass[twocolumn,preprintnumbers,amsmath,amssymb]{revtex4}

\usepackage{graphicx}
\usepackage{dcolumn}
\usepackage{bm}
\usepackage{multirow}
\usepackage{xcolor}
\begin{document}

\title{Steady information transfer through free-space employing orthogonal aberration mode division multiplexing}

\author{Santanu Konwar and Bosanta R Boruah}

 \email{brboruah@iitg.ac.in}
\affiliation{%
Department of Physics,Indian Institute of Technology Guwahati,
Guwahati-781039, Assam, India
}%

\date{\today}

\begin{abstract}
In this paper we propose and experimentally demonstrate information transfer through free-space using a laser beam encoded with multiple orthogonal aberration modes in its phase profile. We use Zernike polynomials which forms a complete set of basis functions to represent the aberration modes. The user information is encoded as amplitudes of multiple Zernike modes whose linear combination is used to define the wavefront of the laser beam in the transmission station. It therefore requires a single phase modulating device since the multiplexing is done digitally providing flexibility over the number and type of modes used. The receiving station uses a wavefront sensor that gives a direct measure of all the aberration modes from a single measurement which is decoded to retrieve the user information. The transmission mechanism enables the use of multiple modes and multiple strengths of each mode to encode the user information and external perturbation compensation, utilizing the completeness of the orthogonal modes. 
\end{abstract}
\maketitle
\section{Introduction}
The last couple of decades have seen rapid developments in free-space optical communication (FSO) systems as an alternative mode of communication that can have advantages in terms of speed and security \cite{chan2006free, khalighi2014survey}. Light beams with helical wavefronts also known as vortex beams or orbital angular momentum (OAM) beams have been proposed for free-space information transfer where the information is encoded as orbital angular momentum states of light \cite{gibson2004free, willner2015optical, padgett2017orbital, willner2017recent}. Multiplexing of OAM modes facilitates enhancing the data transfer rate through free-space up to tera bits per second or more \cite{wang2012terabit, huang2014100}. Capability of the OAM mode for long range communications has also been demonstrated \cite{krenn2014communication, huang2014100}. Perfect vortex beam carrying OAM, where the size of the dark core of the beam is independent of the OAM value, has also been used in free-space communication \cite{zhu2017free}. Not confined to light only, OAM modes in acoustics can also be used \cite{shi2017high} for under water applications. However a major issue of any OAM based free-space communication system is the susceptibility of the mode to even a small amount of atmospheric turbulence \cite{paterson2005atmospheric}. Indeed it was shown that air turbulence may lead to vortex instability with a certain OAM mode after propagation breaking into other OAM modes \cite{lavery2018vortex}. 

Therefore a significant amount of efforts in recent times have been aimed at mitigating the effect of air turbulence in the OAM based FSO systems. Some such efforts are use of modal diversity of different optical modes towards turbulence \cite{PhysRevApplied.10.024020}, use of auto focusing Airy beams \cite{yan2017probability, yan2018controlling}, use of convolutional neural network for adaptive demodulation of the signal at the receiving station \cite{li2018joint}, use of adaptive optics system to compensate for distorted OAM beams \cite{yin2018adaptive} and exploring the shape invariance property of OAM Bessel beams \cite{mphuthi2019free}. Unfortunately most of the turbulence compensation schemes lead to an increase in the processing time at the receiving station \cite{li2018atmospheric} thereby imposing an upper limit on real time decoding.  

Moreover in the quest to achieve higher information carrying capacity, multiplexing of OAM modes is done physically with the help of beam splitters \cite{wang2012terabit, huang2014100}. This limits the flexibility over the maximum number of modes used without disturbing the setup. Besides it has been shown that spatial multiplexing of OAM modes is not an optimal scheme to attain the capacity limits and the same is outperformed by line-of-sight multi-input multi-output transmission and spatial-mode multiplexing \cite{zhao2015capacity}. The root cause behind this limitation can be that the OAM modes alone do not form a complete basis, unless combined with Hermite-Gaussian (HG) modes.

In this paper we propose an orthogonal aberration mode based free-space optical communication system. We use Zernike polynomials \cite{noll1976zernike} which form a complete set of orthogonal basis functions to represent the aberration modes. Multiple Zernike modes whose amplitudes are encoded with user information are multiplexed digitally before transmitting the resulting wavefront using a phase modulating device. The receiving station uses a wavefront sensor to measure the amplitudes of all the Zernike modes constituting the wavefront simultaneously before decoding the user information. The proposed mechanism provides flexibility over the number and types of modes used, and enables compensation of external perturbation effecting the accuracy of the data transmission. In this work we have demonstrated the proposed free-space information transfer in a setup comprising a computer generated holography based transmission and receiving stations.

\section{Generation of single and multiple beams using computer generated holography}
\label{sec2}
Computer generated holography involves computation of the interference pattern between a plane reference wavefront and a user defined object beam wavefront. The computed interference pattern directly or indirectly describes the transmittance of a hologram which is fabricated as a phase plate or implemented using a spatial light modulator. The object beam wavefront can be reconstructed in one of the diffracted beams when a plane wave is incident on the hologram. 
Let us consider that the user defined beam to be generated has the complex amplitude $U(x, y)=e^{i\Phi(x, y)}$ where $(x, y)$ are the coordinates of the hologram plane. The transmittance function of the hologram can be defined as \cite{neil2000wavefront}
\begin{equation}
\label{eq_bin}
t(x, y)=\left\{\begin{array}{ll} 1 & \mbox{if $Real(U) \textgreater 0$ } \\ 0 & \mbox{if $Real(U) \le 0$ } \end{array} \right .
\end{equation}
Here $Real(U)$ represents real part of the function $U$. $t(x, y)$ can define the amplitude transmittance of the hologram in which case the hologram is termed as binary amplitude hologram. A hologram can also be fabricated with $\pi \times t(x, y)$ as the phase delay introduced by the hologram. Such a hologram is termed as binary phase hologram. 

Binary hologram can also be constructed considering a single reference wavefront and a multiple (say, $k$ number of) object beam wavefronts. If $\Phi_j(x, y)$ is the phase in the hologram plane of the $j^{th}$ beam then $U(x, y)=\sum\limits_{j=1}^{j=k} e^{i\Phi_j(x, y)}$. The transmittance function of the binary hologram can still be defined by Eq. \ref{eq_bin}. The hologram to generate a single user defined beam can be termed as singlex hologram while the hologram to generate multiple user defined beams can be termed as multiplex hologram.

In the case of the binary phase or amplitude hologram the plot of $t$ against $\Phi$, at certain location ($x$, $y$), is a square wave \cite{boruah2009dynamic}. Since the Fourier transform of a square wave contains all the odd harmonics, the binary hologram when illuminated by a plane wave results in odd diffraction orders such as $\pm1$, $\pm3$, $\pm5$ and so on. In addition the binary amplitude hologram also results in the undiffracted 0 order. Out of all the diffracted beams, the +1 order beam carries the user defined phase $\Phi$. In order to separate the +1 order beam from the other diffracted beams including the 0 order beam, $\Phi$ should also include a wavefront tilt, $\tau(x, y)=\tau_x x +\tau_y y$, where ($\tau_x, \tau_y)$ are wavefront slopes with respect to ($x$, $y$) axes. Thus $\Phi$ comprises $\phi$ representing phase difference relative to a plane perpendicular to the beam propagation direction and the tilt $\tau(x, y)$ such that $\Phi(x, y)=\phi(x, y)+\tau_x x +\tau_y y$. To construct the multiplex hologram we thus need $k$ number of $(\phi, \tau_x, \tau_y)$ sets. If the light diffracted from the hologram is focused by a lens, the focal spots corresponding to the $\pm1$, $\pm3$, $\pm5$, $\cdots$ are located relative to the zero order at $\tau_c \times$ ($\pm\tau$, $\pm3\tau$, $\pm5\tau$ $\cdots$), where $\tau_c$ is a constant. Thus one can use an iris diaphragm to isolate the +1 order from the other orders and re-collimate the isolated +1 order to recover the user defined phase $\Phi$. The diffraction efficiency of the +1 order beam, using Fourier series analysis, in the case of binary amplitude hologram is found to be $\frac{100}{\pi^2}$\% and in the case of binary phase hologram is found to be $\frac{400}{\pi^2}$\%. The diffraction efficiency in the +1 order beam can be increased to 100\% by constructing a phase hologram with $t(x, y)=Mod(\Phi(x, y), 2\pi)$, where the function $Mod(\Phi(x, y), 2\pi)$ returns the remainder after division of $\Phi$ by $2\pi$. Such a hologram can be termed as blazed grating hologram.

\begin{figure} [!ht]
\centering
\includegraphics [width=8.5 cm] {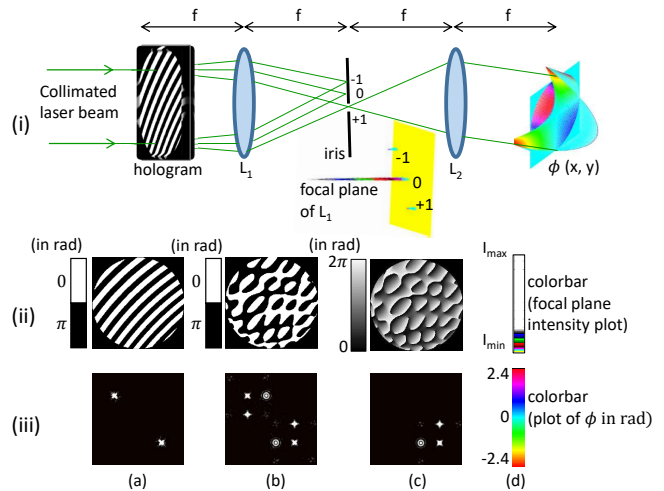}
\caption{Generation of user defined wavefront using computer generated holography: (i) A collimated laser beam is incident 
on a binary amplitude hologram kept at the front focal plane of the lens $L_1$. The diffracted beams from the hologram are focused by the lens $L_1$ and the Fourier transform of the hologram transmittance function is obtained in the back focal plane of $L_1$. The back focal plane thus contains three prominent focal spots comprising $\pm1$ and 0 order beams, as seen in the surface plot. The separation between the $\pm 1$ orders from the 0 order is dependent upon the $\tau(x, y)$ used to construct the binary hologram and is chosen in such a way that an iris diaphragm can separate the +1 order from the rest. The user defined phase profile $\phi$ is realized at the back focal plane of the collimating lens $L_2$. 
(ii) (a$\rightarrow c$) Examples of binary singlex phase, binary multiplex phase and binary multiplex blazed gratings holograms. 
(iii) Representative focal spot patterns in the focal plane of $L_1$ if each of the holograms (ii) 
(a$\rightarrow c$) is placed in the hologram plane of (i). In the case of binary phase holograms both +1 and -1 orders are present while in the case of the blazed grating hologram only the +1 orders are present.}
\label{fig_1}
\end{figure}
Figure \ref{fig_1} depicts the reconstruction of a user defined wavefront (say $\phi(x, y)$) using a binary amplitude hologram and representative focal spot patterns resulting from binary phase and blazed grating holograms.

\section{ Zernike polynomial representation of orthogonal aberration modes}
Zernike polynomials provide a complete set of orthogonal basis functions defined over a unit circle. In this work we use Zernike polynomial $Z_j(r, \theta)$ as described by Noll \cite{noll1976zernike}, where $r$ and $\theta$ are the radial and azimuthal coordinates with $r$ varying from 0 to 1. The Zernike polynomials owing to their description over a circular area provide a convenient way to represent aberrations in optical systems. Indeed a few of the Zernike polynomials represent classical aberrations balanced by lower order aberrations. For instance ($Z_5$, $Z_6$) represent primary astigmatism  ($Z_7$, $Z_8$) represent primary coma balanced by tilt and $Z_{11}$ represent primary spherical aberration balanced by defocus. The expressions of the Zernike polynomials can be converted to Cartesian system using $x=r \cos \theta$ and $y=r \sin \theta$. We therefore can use the computer generated holography technique to generate a laser beam carrying a linear combination of Zernike modes. The hologram is designed using $\phi(x, y)= \sum\limits_j a_j Z_j(x, y)$ where $a_j$ is the co-efficient as well as root mean square amplitude (referred to as $\phi_{RMS}$) of the Zernike mode $Z_j$ in radian. More details on Zernike modes is available in appendix \ref{zernike}.

\section{Multiplexing of Zernike modes to encode the user information}
\label{sec4}
\begin{figure*} [!ht]
\centering
\includegraphics [width=12 cm] {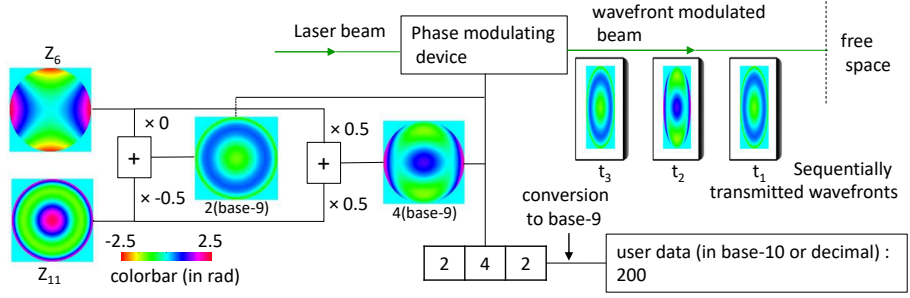}
\caption{Schematic of the encoding scheme in the transmission station: User data equal to decimal number 200 is converted to 242 in base-9 
number system. Linear combinations of two Zernike modes, say $Z_6$ and $Z_{11}$, each with multiple values of $\phi_{RMS}$ (say $\pm$0.5 or 0 radian), are used to map to different digits of base-9 number using the lookup table \ref{tab_1} and the information is sent to a phase modulating device. The device uses the digital data to generate corresponding wavefronts in a beam derived from an incident laser beam. Each wavefront representing a certain base-9 digit is transmitted in a sequential manner through free-space in thee consecutive time slots. Colorbar shows the false colormap in radian of the phase profiles.}
\label{fig_2}
\end{figure*}

\begin{table}
\centering
\caption{An example of lookup table to map the digits of base-9 number system with the linear combinations $\phi_{RMS_1} Z_{6}+\phi_{RMS_2} Z_{11}$ }
\small
\label{tab_1}\
\begin{tabular}{ |p{1cm}|p{0.4cm}|p{0.7cm}|p{0.7cm}|p{0.7cm}|p{0.7cm}|p{0.7cm}|p{0.7cm}|p{0.7cm}|p{0.7cm}| } 
\hline
digits &0&1&2&3&4&5&6&7&8\\
\hline
$\phi_{RMS_1}$&0&0&0&0.5&0.5&0.5&-0.5&-0.5&-0.5\\
\hline
$\phi_{RMS_2}$ &0&0.5&-0.5&0&0.5&-0.5&0&0.5&-0.5\\
\hline
\end{tabular}
\end{table}

In this paper we use one or more number of Zernike modes and multiple $\phi_{RMS}$ of each Zernike mode to describe the phase profile of a laser beam. However when we consider the presence of more than one Zernike mode in the beam then we should take into account the inter-modal cross talk while the modes are measured at the receiving station. We nevertheless can choose combinations of Zernike modes which have the minimal inter-modal cross talk with respect to one another \cite{konwar2018estimation}. If we consider $n$ number of Zernike modes, each having $\phi_{RMS}=a_j$ where $j=1\rightarrow m$, then there will be $m^n$ number of unique linear combinations of $n$ Zernike modes. Thus the user information can be encoded using these $m^n$ unique wavefronts of a laser beam to be transmitted through free-space in a sequential manner. The user data which can be text, numbers, image, etc. is first converted to base-$m^n$ number system (named as base-$m^n$ encoding scheme). Figure \ref{fig_2} depicts an example of the encoding scheme using two Zernike modes each with three value of $\phi_{RMS}$. Thus there are $3^2=9$ number of unique wavefronts each of which can be mapped to the digits of the user data in base-9 number system using a lookup table such as table \ref{tab_1}. One may of course use a different lookup table involving same or different Zernike mode and $\phi_{RMS}$ combinations. More examples of such lookup tables for other encoding schemes are provided in appendix \ref{LUT}).
 It is important to note that unlike OAM based multiplexing, here the multiplexing is done digitally and uses only one phase modulating device to generate the resultant phase profile. Therefore the transmission station has the flexibility over the number of the aberration modes and the number of $\phi_{RMS}$ values for each aberration mode without any modification of the setup. The use of multiple strengths of a single mode to encode the user information, besides the modes forming a complete basis set enhances the information carrying capability of the wavefront. In contrast the information carrying capability of a certain OAM mode is fixed and can not be increased unless other properties such a polarization, wavelength and so on are used for multiplexing. 

\section{Decoding of user information using the type-K modal wavefront sensor}
\label{sec5}
\begin{figure*} [!ht]
\centering
\includegraphics [width=12 cm] {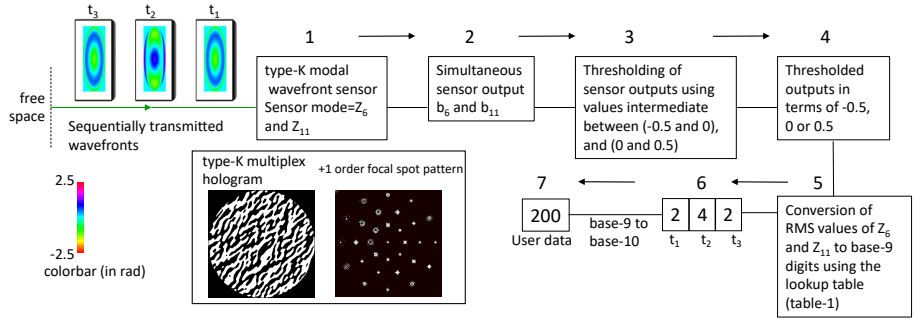}
\caption{Schematic of the decoding scheme at the receiving station: The wavefront from the transmission station is incident on a type-K modal wavefront 
sensor at the receiving station. The receiving station has a priori the key information regarding the Zernike modes used, such as $Z_6$ and $Z_{11}$ and  their strengths, to design an appropriate type-K hologram (such as the one shown in the inset). For a given incident wavefront the type-K hologram generates a focal spot pattern, such as the one shown in the inset. The focal spot pattern is employed to obtain the sensor outputs for all the Zernike modes used during encoding. The type-K sensor outputs, such as $b_6$ and $b_{11}$, are first thresholded and then the $\phi_{RMS}$ values after thresholding ( such as 0, -0.5 or 0.5 radian)  are converted to appropriate base-n digit using the lookup table (such as table \ref{tab_1} for base-9). Thus the receiving station a priory needs to have the lookup table. Each wavefront incident on the type-K sensor results in one digit of the base-n number system and base-n digits corresponding to all the wavefronts received at different time slots are converted to the user data.}
\label{fig_3}
\end{figure*}
The laser beam with the user information encoded wavefront travels the free space and is then incident on a modal wavefront sensor at the receiving station. The modal wavefront sensor should be able to measure simultaneously the $\phi_{RMS}$ of all the Zernike modes present over a range of amplitudes. However the conventional modal wavefront sensors \cite{neil2000new, booth2003direct} can measure only small aberration amplitudes and suffer from inter-modal cross talk when multiple aberrations are to be measured. If large aberration amplitudes are to be measured with the same accuracy then the sensing operation is carried out in a loop thereby lowering the sensing frame rate \cite{Neil:00, dong2012response}. In our work we use a recently proposed type-K modal wavefront sensor \cite{Konwar:19} to measure the multiple Zernike modes (referred to as sensor modes) present in the incident wavefront. The type-K sensor estimates the aberrations from the intensity data captured in a single camera image. It has a large range of linear response that can be more the ten times the linear response range of a conventional modal wavefront sensor and $\phi_{RMS}$ of two or more sensor modes can be measured with reduced inter-modal cross talk. The outputs of the type-K sensor (denoted as $S_K$) representing $\phi_{RMS}$ in radian of the sensor modes undergo a thresholding process to take into account any minor deviation between the $\phi_{RMS}$ used during encoding and $\phi_{RMS}$ as estimated. For instance if the $\phi_{RMS}$ of a Zernike mode during encoding is -0.5 radian, 0 or 0.5 radian, the estimated $\phi_{RMS}$ is compared with some intermediate value between -0.5 and 0 or 0 and 0.5, say -0.25 or 0.25. If the estimated $\phi_{RMS}$ is 0.4 radian, it becomes 0.5 radian after thresholding. The thresholded $\phi_{RMS}$ of all the sensor modes are then converted to a digit of the corresponding number system using the appropriate lookup table. Figure \ref{fig_3} presents the schematic of the decoding process for the encoded wavefront shown in Fig. \ref{fig_2}.

The type-K modal wavefront sensor comprises primarily a multiplex hologram called the type-K hologram and a camera. The hologram generates a set of +1 order beams for each sensor mode for a given incident wavefront. The hologram effectively adds or subtracts specific amount of the sensor mode to the phase profiles of these +1 order beams. The central intensities of the set of +1 order focal spots are then used to calculate the type-K sensor output $S_K$. Figure \ref{fig_3} inset shows a representative type-K hologram and the resulting +1 order focal spots pattern. The focal spots pattern comprises 30 focal spots, arranged in concentric circles, for three sensor modes with 10 focal spots to estimate each sensor mode.  There is one more focal spot at the centre that corresponds to a +1 order beam without any addition or subtraction of aberration by the hologram. More details of the type-K sensor is available in appendix \ref{type-K}.

It is to be noted that for correct decoding of the user information the receiving station should have two key information, one is the number and types of aberration modes used for encoding (required to design the type-K hologram) and the other is the lookup table (required to connect the sensor outputs to the appropriate base-n digit). As the beam propagates through the free space, it undergoes certain amount of divergence, depending on the distance traveled and to some extent on the type of the orthogonal mode used. Therefore if the same size of the type-K hologram is to be used for all the distances traveled by the beam, the incident beam is first to be demagnified. Besides the type-K sensor output may require to be scaled (normalized) before the same is used for decoding the user data.

 \section{Experimental implementation}
\begin{figure*} [!ht]
\centering
\includegraphics [width=12 cm] {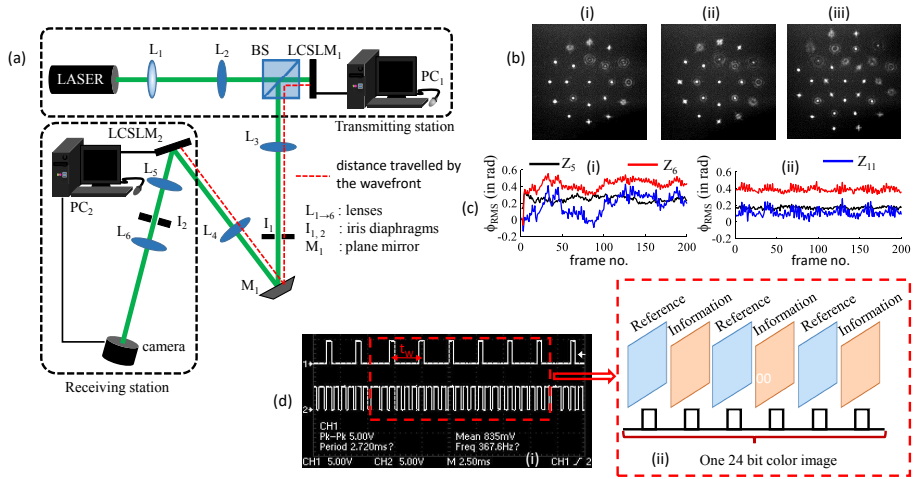}
\caption{Schematic of the free-space optical communication system: (a) A collimated laser beam is incident on a reflective type ferroelectric liquid crystal spatial light modulator (LCSLM$_1$). The device can display binary holograms sent by PC$_1$ at a rate of 1440 Hz. The user information is encoded as $\phi_{RMS}$ of Zernike modes present in the +1 order beam diffracted from 
LCSLM$_1$ (as described in section \ref{sec2}). The beam after traveling the free-space is incident on another reflective liquid crystal spatial light modulator (LCSLM$_2$) that displays an appropriate type-K multiplex hologram via PC$_2$. The resulting +1 order focal spots are captured by a digital camera and the image is read by a computer program in  PC$_2$ that decodes the user information (as described in section \ref{sec5}). The distance in free-space traveled by the beam is 2.4 metres. In this implementation the LCSLM$_1$ and LCSLM$_2$, although not essentially required, are arranged to be on conjugate planes with respect to one another. More details of the experimental implementation is found in appendix \ref{methods}.
(b) Experimental focal spot patterns recorded by the camera with type-K hologram for (i) base-9 encoding scheme (using sensor modes $Z_6$ and $Z_{11}$ ), (ii) base-25 encoding scheme (using sensor modes $Z_5$ and $Z_{11}$), and (iii) base-27 encoding scheme (using sensor modes $Z_5$, $Z_6$ and $Z_{11}$). In all the three cases the incident beam has a plane wavefront. 
(c) Experimental type-K sensor outputs over several camera frames, in the presence of turbulence introduced by a table fan, (i) before external perturbation compensation and (ii) after external perturbation compensation. 
(d) (i) Screen shot of the
oscilloscope showing the display timings of binary holograms at 1440 Hz LCSLM$_1$ and the timings of user wavefront generation at an interval of $t_w$.
(d) (ii) Schematic showing the sequence of reference wavefront and information carrying wavefront from the transmission station, over a duration equal to the display time of a 24 bit color image by the LCSLM$_1$. 
}
\label{fig_4}
\end{figure*}
A schematic of the experimental arrangement comprising the transmission station based on a ferroelectric LCSLM (LCSLM$_1$) acting as the binary hologram and a receiving station based on a type-K hologram written on a nematic LCSLM (LCSLM$_2$), separated by a distance of 2.4 metres, is seen in Fig. \ref{fig_4} (a). Figure \ref{fig_4} (b) shows some experimental focal spot patterns resulting from the type-K hologram. Instead of implementing a hologram to realise wavefront modulation in the transmission station, one can also use direct phase modulation by a phase modulating LCSLM or a deformable membrane mirror. The latter may provide wavefront modulation at frame rate upto tens of KHz. Besides the type-K hologram instead being implemented using an LCSLM can also be fabricated as a phase plate. The decoding speed can be enhanced significantly by using a set of point photo detectors to record the central intensities of the focal spots instead of the camera. 

\section{External perturbations compensation utilizing the completeness of the Zernike modes}
We first assess the consistency of accurate decoding of user data over  a period of time. We encode the wavefront in the transmission station using three Zernike modes, $Z_5$, $Z_6$ and $Z_{11}$ and estimate the strengths of the same three modes in the receiving station using the type-K sensor. We notice that owing to the movement and externally caused perturbations in the information carrying beam, the values of $\phi_{RMS}$ at the receiving station fluctuates to some extent over time even though there is no change in the aberration strengths from the transmission station. We use a table fan to create instability in the air in between the two stations so as to enhance the fluctuations in $S_K$. The table fan creates a circulating turbulent air with an average speed between 4 m/s to 5 m/s. Figure \ref{fig_4} (c) (i) shows such an induced case of sensor output variation over different measurement frames when a fixed combination of $\phi_{RMS}$=(0, 0.5, 0) of ($Z_5$, $Z_6$, $Z_{11}$) is transmitted. Figure \ref{fig_4} (d) (i) shows the binary pattern display timing of the LCSLM$_1$ at the rate of 1440 Hz. However due to limited frame rate of the camera we display binary holograms at the rate 360 Hz only with an interval of $t_w$ as seen in Fig. \ref{fig_4} (d) (i). In order to make $S_K$ steady against the perturbations we modify the transmission scheme of the encoded wavefronts, such that each information carrying wavefront is preceded by a reference plane wavefront as depicted in Fig. \ref{fig_4} (d) (ii). The reference wavefront when incident on the type-K hologram results in a reference focal spot pattern whose central focal spot location gives a measure of beam movement. Since the Zernike modes form a complete set of basis functions, any arbitrary change in the reference phase profile caused by the perturbation can be expressed as a linear combination of a subset of the Zernike modes. Therefore primary effect of perturbation other than the beam movement can be estimated from a measure of the type-K sensor outputs corresponding to the reference wavefront. Since type-K sensor has a linear response range upto several radian of $\phi_{RMS}$ for each sensor mode, corrected $S_K$ for each information frame can be obtained, after beam movement compensation, as $S_K$ (information frame)-$S_K$ (preceding reference frame). More details about the perturbation compensation scheme is available in appendix \ref{purtcomp}.
On application of the perturbation compensation, the type-K sensor in the receiving station provides a much more consistent measure of the sensor modes as indicated by the plot in Figure 4 (c) (ii). The compensation scheme to be effective the nature of the external perturbation should not vary between the reference wavefront and the following information wavefront.

\section{Results and discussion}
\begin{figure} [!ht]
\centering
\includegraphics [width=8.5 cm] {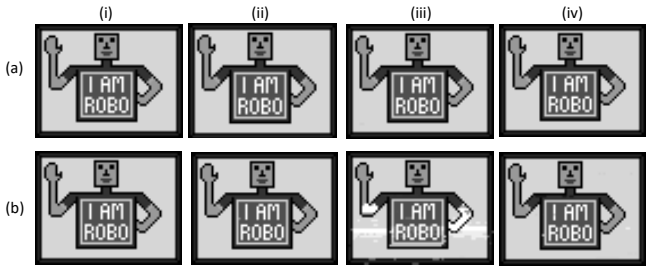}
\caption{(a) (i) Original image sent by the transmission station. (a) (ii, iii, iv) Images (experimental) as detected with external perturbation compensation when 
the user data is transmitted using base-3, base-5 and base-9 encoding schemes. (b) (i, ii) Images (experimental) as detected with external perturbation compensation when 
the user data is transmitted using base-25 and base-27 encoding schemes.(b) Images (experimental) as detected (iii) without external perturbation compensation and (iv) with external perturbation compensation, when the user data is transmitted through turbulent air created by a table fan and using base-25 encoding scheme.}
\label{fig_5}
\end{figure}
We choose the three Zernike mode combinations ($Z_5$, $Z_6$, $Z_{11}$) to encode the user data. In our first experiment we transmit a gray scale image of size 65$\times$50 pixels with each pixel as an 8 bit integer. The image pixels are encoded using base-3, 5, 9, 25 and 27 encoding schemes. At the receiving station the wavefronts are decoded both without perturbation compensation and with perturbation compensation. It can be seen in appendix \ref{LUT} that base-3 and 5 encoding schemes use a single Zernike mode while base-9 and 25 encoding schemes use two Zernike modes. However due to use of more number of strengths for each mode, the information content of $m$ number of wavefronts in the case of base-5 scheme has increased by $\frac{5^m-1}{3^m-1}\times 100$\% relative to base-3 scheme and the same for base-25 has increased by $\frac{25^m-1}{9^m-1}\times 100$\% relative to base-9 scheme. In comparison with OAM mode based data transfer, the maximum number of unique information carried by a beam, employing two Zernike modes multiplexing using the base-25 scheme, over 8 modulation cycles is $25^8-1=1.5259\times 10^{11}$ which in the case of two OAM modes multiplexing is $4^8$-1=65535. The effective rate of data transfer in our proposed scheme is decided by both the encoding scheme and minimum interval (i.e. time period of one modulation cycle) at which the user defined wavefront can be shaped by the phase modulating device. Figure 5 (a) (i) shows the original image transmitted and Figs. \ref{fig_5} (a) (ii), (iii), (iv) and Figs. \ref{fig_5} (b) (i) and (ii) show the detected images with perturbation compensation when the data is transmitted using different encoding schemes. It is noticed that in the case of base-3 and base-9 encoding schemes the image is detected without a single wrong pixel. Although in the other cases there is an error of maximum 1 to 4 pixels, even though there is still no major observable defects in the detected images. In order to further increase the external perturbation we use the table fan to introduce turbulence in the medium. The directly detected image transmitted as base-25 numbers which travel through the turbulent air is seen in Fig. \ref{fig_5} (b) (iii) showing several wrong pixels. However on application of perturbation compensation the number of wrong pixels reduces drastically as seen in Fig.  \ref{fig_5} (b) (iv).

\begin{table}
\centering
\caption{Table showing  the total number ($N_T$) of transmitted 8 bit integers and alphabets sent using base-3, 5, 9, 25 and 27 encoding schemes against the corresponding numbers of wrongly detected data without the perturbation compensation, $N_W$(npc) and with the perturbation compensation, $N_W$(wpc).}
\small
\label{tab_2}\
\begin{tabular}{ |c|c|c|c|c|c|c| } 
\hline
\multirow{3}{*}{\textbf{Base-n}} &  \multicolumn{3}{c|}{\textbf{8 bit integers}} & \multicolumn{3}{c|}{\textbf{alphabets}}\\
\cline{2-7}
&$N_T$ & $N_W$(npc) & $N_W$(wpc)  &$N_T$ & $N_W$(npc) & $N_W$(wpc)    \\

\hline
Base-3
&262&0&0&363&1&0 \\ \hline
Base-5
&348&0&0&363&0&0 \\ \hline
Base-9
&393&17&2&363&0&0 \\ \hline
Base-25
&696&333&3&726&16&3 \\ \hline
Base-27
&917&68&2&726&31&0 \\ \hline
\end{tabular}
\end{table}

We then perform yet another experiment to transmit a set of 8 bit integers and alphabets in a similar manner as the above experiment. The results presented in table \ref{tab_2} confirm that when the user data at the receiving station is decoded with perturbation compensation, the effect of perturbation significantly gets eliminated. Accuracy of the transmission increases when the data is encoded as base-3 and base-5 numbers. 

The relatively small number of wrong detections of user data in some of the schemes is attributed to the low modulation frequency of 360 Hz owing to the limitation of the camera used in the present setup. Even a three fold increase in the modulation frequency should make the perturbation compensation much more effective so as to eliminate wrong detection of user data altogether in all the schemes. Under extreme cases of turbulence, use of an encoding scheme with a smaller base such as base-3 and choosing larger separation between $\Phi_{RMS}$ values, such as 0, $\pm 1$ instead of 0, $\pm 0.5$, will further improve the perturbation compensation.

\section{Summary}
We have demonstrated free-space information transfer using a laser beam whose wavefront is encoded with a linear combination of Zernike modes which form a complete set of orthogonal basis. User information is converted to amplitudes of one or more Zernike modes using multiple amplitude values for each Zernike mode. Multiplexing of modes is done digitally and a single phase modulating device is used to transmit the data as sequentially transmitted wavefronts. Use of a complete set of orthogonal functions and multiple amplitudes of a single mode, to encode the wavefront, opens the door towards reaching optimum limit of information content in the beam. The receiving  station uses a type-K modal wavefront sensor that provides amplitudes of all the Zernike modes present in the incident wavefront from just one measurement using a camera. The use of the type-K sensor enables measurement of both small as well as large amplitudes of each Zernike mode with the same accuracy. We have also demonstrated a scheme to compensate for external perturbations such as turbulence in the air so as to make the information transfer steady. The proposed information transfer scheme has a built in security feature against eavesdropping since the receiving station must a priori know the number, types and amplitudes of the Zernike modes used and the lookup table mapping the Zernike mode amplitudes with the user data. Our experimental arrangement however can be considered as a proof-of-principle setup only and the same was not optimized in the terms of speed, accuracy, information content in the wavefront, and the distance traveled in free-space. There is nevertheless an equivalent scope in the proposed scheme, similar to the OAM mode based systems, to further enhance the information content of the beam by incorporating polarization and wavelength division multiplexing over and above the Zernike mode division multiplexing. The scheme introduced in this paper will apply to other types of orthogonal modes forming a complete basis set, in addition to the Zernike modes, which was used in this work.

\bigskip
The authors wish to acknowledge the financial support from the Department of Electronics and Information Technology (DeitY), India, vide its letter no 12(4)/2011-PDD.

\appendix
\section{Methods}
\label{methods}
\subsection{Construction of binary hologram in the transmission station}
Binary hologram to generate the +1 order beam carrying user defined phase profile $\phi(x, y)$ with $(x, y)=-1\rightarrow+1$ is constructed over 480$\times$480 pixels using $\tau_x= 165\pi$ and $\tau_y= -41\pi$. An appropriate linear combination of Zernike modes $Z_j(x, y)$ constitutes $\phi(x, y)$ and $t(x, y)$ is computed using $U(x, y)=e^{\phi(x, y)+\tau_x x+\tau_y y}$ in Eq. \ref{eq_bin}.  $t(x, y)$ is multiplied by a unit circle centred at $(x, y)=(0, 0)$ and is then converted to a binary image to be displayed on the ferroelectric LCSLM (ForthDD, SXGA-R3) which has a pixel pitch = 13.62 $\mu$m. To be noted that the +1 order beam from the LCSLM has some residual aberrations which may be corrected holographically in order to get a near ideal +1 order beam.

\subsection{Incident laser beam} The beam incident on the binary hologram on the ferroelectric LCSLM is derived from a He-Ne Laser (wavelength = 632.8 nm, output power = 10 mW). Experiments are performed at $\approx$ 50\% of the laser power. The +1 order beam carrying the user information has a diameter of $\approx$6.5 mm.

\subsection{Construction of type-K hologram and recording of the focal spot pattern} The type-K multiplex binary holograms are constructed over 750$\times$750 pixels and are displayed as binary images on a nematic LCSLM (HoloEye  LC-R-1080) that has a pixel pitch = 8.1 $\mu$m.
$t(x, y)$ is computed using $k$ sets of ($\phi(x, y)$, $\tau_x$, $\tau_y$), where $k$=11 for base-3 and 5 encoding schemes, $k$=21 for base-9 and 25 encoding schemes and $k$=31 for base-27 encoding scheme. The +1 order beam at the centre of the type-K focal spot pattern has $\tau_x=153.8\pi$ and  $\tau_y= 100.9\pi$. For the base-3 encoding scheme the type-K hologram generates 5 pairs of bias beams with $a_v$=(-1 -0.5 0 0.5 1) radian to detect a single sensor mode. These beams are directed to form a circle using relative tilt with respect to the central beam (defined as $\Delta \tau$=$\sqrt{\Delta \tau_x^2+ \Delta \tau_y^2}$) as 21$\pi$. For the base-5 encoding scheme the type-K hologram generates 5 pairs of bias beams with $a_v$=(-1.5 -0.75 0 0.75 1.5) radian to detect a single sensor mode. These beams are also directed to form a circle using $\Delta \tau$=21$\pi$. Type-K holograms for base-9 and base-25 encoding schemes are designed for two sensor modes using $a_v$=(-1 -0.5 0 0.5 1) and $a_v$=(-1.5 -0.75 0 0.75 1.5) for each sensor mode, respectively. For both encoding schemes bias beams are directed to form two concentric circles using $\Delta \tau$=$14\pi$ and $24.5\pi$. Type-K hologram for base-27 encoding scheme is designed for three sensor modes using $a_v$=(-1 -0.5 0 0.5 1) for each sensor mode and the bias beams are directed to form three concentric circles using $\Delta \tau$=11.5$\pi$, $21\pi$ and $30\pi$. The bias beams for each type-K hologram use $b$=0.7 radian.

The focal spot pattern is recorded by a CMOS camera (Basler, A504K) having pixel resolution = 1280$\times$1024, pixel pitch = 12 $\mu$m, bit depth=8 bit and a full frame rate= 500 fps.

\section{Zernike polynomials}
\label{zernike}
Single index Zernike polynomial $Z_j(r, \theta)$ as described by Noll \cite{noll1976zernike} can be written as
\begin{equation}
\begin{array}{l}
{\left. \begin{array}{l}
Z_{even\, j}=\sqrt{n+1} R_n^m (r) \sqrt{2} \cos(m\theta)\\
Z_{odd\, j}=\sqrt{n+1} R_n^m (r) \sqrt{2} \sin(m\theta)
\end{array} \right\}\,\mbox{for m$\ne$0}}\\
Z_{j}=\sqrt{n+1} R_n^m (r)\,\mbox{for m=0}
\end{array}
\end{equation}
where
\begin{equation}
R_n^m(r)=\sum_{q=0}^\frac{(n-m)}{2} \frac{(-1)^q (n-q)!}{q! \left[\frac{n+m}{2}-q\right]!\left[\frac{n-m}{2}-q\right]!} r^{n-2q}
\end{equation}
Here $r$ and $\theta$ are the radial and azimuthal coordinates describing the circular area with $r$ varying from 0 to 1. 
$n$ and $m$ are integers and satisfy $m\le n$, $n-|m|$ = even. The index $j$ represents mode ordering and is a function of $n$ and $m$. 
The expressions of the Zernike polynomials can be converted to Cartesian system using ($x=r \cos \theta$, $y=r \sin \theta$). Expressions of a few Zernike modes $Z_j(x, y)$ 
representing some low order optical aberrations used in this paper are given below

 \begin{table}[!ht]
\label{tab:zernike}
\begin{center}
\begin{tabular}{llll}
\hline
j &  Z$_j$(x, y) & Name\\
\hline
5 & $2\sqrt{6}xy$ & Astigmatism at $\pm45^\circ$\\
6 &  $\sqrt{6}(x^2-y^2)$ & Astigmatism at 0 or 90$^\circ$\\
7 &$\sqrt{8} (3(x^2+y^2)-2)y$ & Primary y coma\\
8 & $\sqrt{8} (3(x^2+y^2)-2)x$ &  Primary x coma\\
11 &$ \sqrt{5} (6(x^2+y^2)^2-6(x^2+y^2)$& Primary spherical\\
&+1)& aberration\\
\hline
\end{tabular}
\end{center}
\end{table}

\section{Lookup tables for different encoding schemes}
\label{LUT}
In the transmission station the user information is first converted to base-n number system, where we use n=3, 5, 9, 25 and 27.  A number system with base-n will comprise n digits 
such as 0, 1, 2,$\cdots$, (n-1). We use multiple $\phi_{RMS}$ (RMS amplitudes) values of one or more Zernike modes to represent each digit of a  base-n number. 
In the lookup table \ref{LUtable1} we show an example of how the amplitude ($\phi_{RMS_1}$) of a single Zernike mode $Z_{i1}$ can be mapped to the digits of base-3 and base-5 
numbers and how the amplitudes ($\phi_{RMS_1}$ and $\phi_{RMS_2}$) of two Zernike mode $Z_{i1}$ and $Z_{i2}$ can be mapped to the digits of a base-9 number. The two tables thus represent base-3, 5, and 9 encoding schemes.
\begin{table}[!ht]

\centering
 \caption{Lookup tables mapping $\phi_{RMS_1}$ and $\phi_{RMS_2}$ to the digits of base-3, 5 and 9 number systems. Base-3 and 5 encoding schemes use one Zernike mode while base-9 encoding scheme uses two Zernike modes.}
\begin{tabular}{ ll } 
\begin{tabular}{ |c|c|c| } 
\hline
Base-n & digit & $\Phi_{RMS_1}$\\

\hline
\multirow{3}{*}{Base-3}
&0 &0 \\
&1 &0.5 \\ 
&2 &-0.5 \\ \hline
     
\multirow{6}{*}{Base-5}
&0 &0 \\
&1 &0.5 \\ 
&2 &-0.5 \\
&3 &1 \\ 
&4 &-1 \\ 
&&\\
\hline
\end{tabular}
         
&

\begin{tabular}{ |c|c|c|c| } 
\hline
Base-n & digit & $\phi_{RMS_1}$& $\phi_{RMS_2}$\\
\hline
\multirow{9}{*}{Base-9}
&0 &0 & 0\\
&1 &0 & 0.5\\ 
&2 &0 & -0.5\\
&3 &0.5 & 0\\
&4 &0.5 & 0.5\\ 
&5 &0.5 & -0.5\\
&6 &-0.5 & 0\\
&7 &-0.5 & 0.5\\ 
&8 &-0.5 & -0.5\\ \hline
\end{tabular}
\end{tabular}
\label{LUtable1}
\end{table}
In a similar manner the lookup tables \ref{LUtable2} we show an example of how the amplitude ($\phi_{RMS_1}$ and $\phi_{RMS_2}$) of two Zernike mode $Z_{i1}$ and $Z_{i2}$ can 
be mapped to the digits of a base-25 number and how the amplitudes ($\phi_{RMS_1}$, $\phi_{RMS_2}$ and $\phi_{RMS_3}$) of three Zernike mode $Z_{i1}$, $Z_{i2}$ and $Z_{i3}$ can 
be mapped to the digits of a base-27 number. The two tables represent base-25 and 27 encoding schemes.

\begin{table*}[!ht]
\centering
 \caption{Lookup tables mapping $\Phi_{RMS_1}$, $\Phi_{RMS_2}$ and  $\Phi_{RMS_2}$ to the digits of base-25 and 27 number systems. Base-25 encoding scheme uses two Zernike mode modes while base-27 encoding scheme uses three Zernike modes.}
\begin{tabular}{ ll } 
\begin{tabular}{ |c|c|c|c| } 
\hline
Base-n & digit & $\Phi_{RMS_1}$& $\Phi_{RMS_2}$\\
\hline
\multirow{27}{*}{Base-25}
&0 &0 & 0\\
&1 &0 & 0.5\\ 
&2 &0 & -0.5\\
&3 &0 & 1\\
&4 &0 & -1\\
&5 &0.5 & 0\\
&6 &0.5 & 0.5\\ 
&7 &0.5 & -0.5\\
&8 &0.5 & 1\\ 
&9 &0.5 & -1\\
&10 &-0.5 & 0\\
&11 &-0.5 & 0.5\\ 
&12 &-0.5 & -0.5\\
&13 &-0.5 & 1\\ 
&14 &-0.5 & -1\\
&15 &1 & 0\\
&16 &1 & 0.5\\ 
&17 &1 & -0.5\\
&18 &1 & 1\\ 
&19 &1 & -1\\
&20 &-1 & 0\\
&21 &-1 & 0.5\\ 
&22 &-1 & -0.5\\
&23 &-1 & 1\\ 
&24 &-1 & -1\\ 
&&&\\
&&&\\ \hline
\end{tabular}
&
\hspace{1cm}
\begin{tabular}{ |c|c|c|c|c| } 
\hline
Base-n & digit & $\Phi_{RMS_1}$& $\Phi_{RMS_2}$& $\Phi_{RMS_3}$\\
\hline
\multirow{27}{*}{Base-27}
&0 &0&0 & 0\\
&1 &0&0 & 0.5\\ 
&2 &0&0 & -0.5\\
&3 &0&0.5 & 0\\
&4 &0&0.5 & 0.5\\ 
&5 &0&0.5 & -0.5\\
&6 &0&-0.5 & 0\\
&7 &0&-0.5 & 0.5\\ 
&8 &0&-0.5 & -0.5\\
&9 &0.5 &0& 0\\
&10 &0.5 &0& 0.5\\ 
&11 &0.5 &0& -0.5\\
&12 &0.5 &0.5 & 0\\
&13 &0.5 &0.5& 0.5\\ 
&14 &0.5 &0.5& -0.5\\
&15 &0.5 &-0.5 & 0\\
&16 &0.5 &-0.5& 0.5\\ 
&17 &0.5 &-0.5& -0.5\\
&18 &-0.5 &0 & 0\\
&19 &-0.5 &0& 0.5\\ 
&20 &-0.5 &0& -0.5\\
&21 &-0.5 &0.5 & 0\\
&22 &-0.5 &0.5& 0.5\\ 
&23 &-0.5 &0.5& -0.5\\
&24 &-0.5 &-0.5 & 0\\
&25 &-0.5 &-0.5& 0.5\\ 
&26 &-0.5 &-0.5& -0.5\\
\hline
\end{tabular}
\end{tabular}
\label{LUtable2} 
\end{table*}

Phase profile of the encoded wavefront can be expressed as $\phi(x, y)=\phi_{RMS_1} Z_{i1}$ for base-3 and base-5 encoding schemes, as 
$\phi(x, y)=\phi_{RMS_1} Z_{i1}+\phi_{RMS_2} Z_{i2}$ for base-9 and base-25 encoding schemes, and $\phi(x, y)=\phi_{RMS_1} Z_{i1}+\phi_{RMS_2} Z_{i2}+\phi_{RMS_3} Z_{i3}$ for base-27 encoding schemes. As stated the above lookup tables are examples only and the proposed free-space transmission mechanism is applicable for other combinations of RMS amplitudes of orthogonal aberration modes as well. For instance multiplexing of 3 Zernike modes each with 5 strengths will result in base-125 encoding scheme. In so far as information capacity of the encoded beam is concerned, $m$ wavefronts encoded using say base-n encoding scheme can define a maximum decimal number equal to $n^m-1$ and thus can carry equal number of unique user informations. Therefore enhancement in the information capacity over $m$ modulation cycles in base-n scheme relative to base-n$^\prime$ scheme is  $\frac{n^m-1}{{n^\prime}^m-1}\times 100$\%.

\section{Type-K modal wavefront sensor}
\label{type-K}
\begin{figure*} [!ht]
\centering
\includegraphics [width=16 cm] {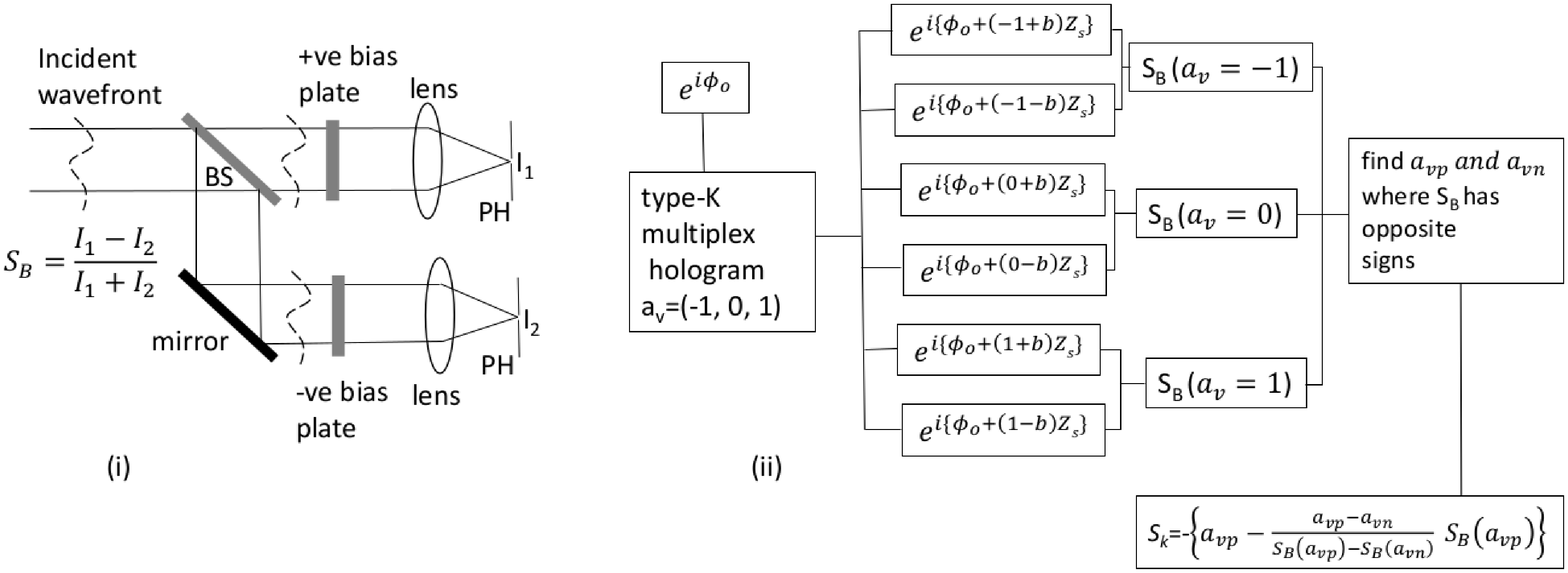}
\caption{(i) Working principle of bias beam based modal wavefront sensor, (ii) the type-K modal wavefront sensing scheme for one 
sensor mode using three pairs of bias beams with $a_v$=(-1, 0, 1).}
\label{fig_sp1}
\end{figure*}

A modal wavefront sensor provides the strength of an aberration mode such a Zernike mode present in the wavefront of a laser beam directly without much processing unlike the 
zonal wavefront sensor. The zonal wavefront sensor uses a reference wavefront and post acquisition data processing to provide strengths of the Zernike modes present in the beam. 
The magnitude of a Zernike mode present in a beam can be directly measured in the bias beam based modal wavefront sensor proposed by Neil et al. \cite{neil2000new}. The principle of the 
bias beam based modal wavefront sensor is depicted in Fig. \ref{fig_sp1} (i). The laser beam whose wavefront contains a Zernike mode $Z_s$ to be sensed (referred to as the sensor 
mode) is first divided into two beams using a beam splitter, each having the identical wavefront and equal intensity. One of the beams is added with $\phi_{RMS}=b$ amount of the 
sensor mode using a phase plate called positive (+ve) bias plate and the same amount of the sensor mode is subtracted from the other beam using another phase plate called 
negative (-ve) bias plate. The beam emerging from the +ve bias plate is called +ve bias beam and the beam emerging from the -ve bias plate is called -ve bias beam. Therefore if 
the incident beam has the complex amplitude $e^{i\phi_o}$, the complex amplitudes of the +ve bias beam is $e^{i\phi_o+b Z_s}$ and the -ve bias beam is $e^{i\phi_o-b Z_s}$. 
The two beams are focused onto two pinholes (PH) and two photodetectors are used to record the central intensities $I_1$ and $I_2$ of the two focal spots. Neil et al. showed that 
the difference $I_1-I_2$ gives a measure of the Zernike mode $Z_s$ present in the beam. Thus a modal wavefront sensor can be designed such that its sensor out $S_A=I_1-I_2$ 
(referred to as type-A sensor output). The sensitivity of the sensor output reaches it maximum when $b\approx$0.7 radian. The sensor output expression can be modified \cite{booth2003direct}
as $S_B=\frac{I_1-I_2}{I_1+I_2}$ in which case it is called type-B sensor output. However the type-A or the type-B modal wavefront sensors provide linear response upto about 
$\pm$0.4 radian of the sensor mode and they are very much effected by modes other than the sensor mode present in the beam which is known as inter-modal cross talk. 
These two sensors are therefore effective for a weakly aberrated incident beam only, if the sensor output is obtained from a single measurement of $I_1$ and $I_2$.

Type-A or type-B sensors can also be implemented holographically. The $t(x, y)$ of the hologram is computed to results in two +1 order beams, one using $\phi=b Z_s$ and the other 
using $\phi=-b Z_s$. For a beam with complex amplitude $e^{i\phi_o}$ incident on the hologram, the complex amplitudes of the two +1 orders will be $e^{i\phi_o+b Z_s}$ and 
$e^{i\phi_o-b Z_s}$. Thus the two +1 order beams effectively behave as the +ve bias and the -ve bias beams, which can be focused using single lens and type-A or type-B sensor 
outputs can be calculated. Such a sensor is called holographic modal wavefront sensor. 

If we consider that we incorporate a variable amount $a_v$ of the sensor mode into the corresponding bias beams then their complex amplitudes are $e^{i\phi_o+(a_v+b) Z_s}$ 
and $e^{i\phi_o+(a_v-b) Z_s}$. The type-B sensor output using these two bias beams can be termed as $S_B(a_v)$. It can be shown \cite{Konwar:19} that the negative of the root of 
the equation  ${S_B|}_{a_v}=0$ corresponds to $\phi_{RMS}$ of the sensor mode, which to a large extent is independent of the presence of other aberration modes and is valid 
irrespective of the range of the sensor mode strength. This is primarily because near the root of the equation ${S_B|}_{a_v}=0$, the bias beams have minimal aberrations 
approximately equal to the bias aberrations only. The type-K sensor is designed with the sensor output defined as $S_K=-a_{v0}$, such that ${S_B|}_{a_v=a_{v0}}=0$. In the case of the 
type-K sensing scheme for each sensor mode to be measured a certain number ($k_s$) of pairs of bias beams are generated using a multiplex hologram. Each bias pair has an additional 
$a_v$ amount of the sensor mode. For $k_s$=3 we may take $a_v$=(-1, 0, 1). Central intensities of the $k_s$ pairs of bias beams are used to calculate $S_B(a_v)$. A fairly accurate 
estimation of $a_{v0}$ is obtained by doing a linear interpolation between $S_B(a_{vp})$ and $S_B(a_{vn})$, where $a_{vp}$ and $a_{vn}$ are nearest $a_v$ values on the two sides of 
the root. Therefore 
\begin{equation}
\label{eq_1}
S_K=-\left\{(a_{vp})-\frac{a_{vp}-a_{vn}}{S_B(a_{vp})-S_B(a_{vn})}S_B(a_{vp}) \right\}
\end{equation}
 The multiplex hologram to generate $k_s$ pairs of bias beams is called the type-K hologram. Instead of just one sensor mode the type-K hologram can also be designed for more than 
 one sensor mode. The range of linear response of the sensor is decided by the minimum and the maximum $a_v$ values and the range can hence be enhanced by choosing $a_v$ over a 
 large range. The accuracy of the sensor output is decided by the separation between the two $a_v$ values used in Eq. \ref{eq_1}. The hologram is designed to give rise to a 
 concentric focal spot pattern for the convenience of recording using a single camera frame. Besides the type-K hologram may also incorporate a reference +1 order beam whose 
 position can be used as the centre of the focal spot pattern. Figure \ref{fig_sp1} (ii) depicts the type-K modal sensing scheme for an incident complex amplitude $e^{i\phi_o}$ to detect a single sensor mode using $a_v$=(-1, 0, 1). It is to be noted that the same type-K hologram can detect sensor mode strengths of both positive and negative polarity. In contrast OAM modes with opposite polarity will require two different holograms to detect the two modes. Therefore change in polarity of the Zernike mode amplitude is not equivalent to the change in polarity of an OAM mode as in the latter case they constitute two different modes.

\section{External perturbations compensation utilizing the completeness of the Zernike modes}
\label{purtcomp}
\begin{figure} [!ht]
\centering
\includegraphics [width=8 cm] {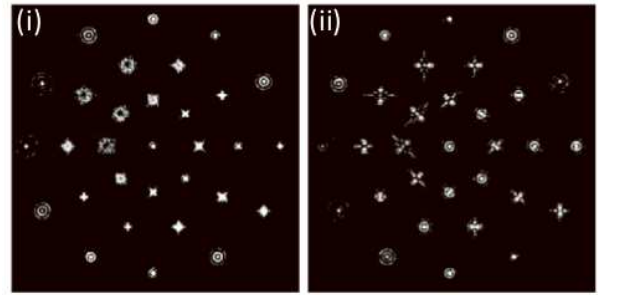}
\caption{A representative focal spot pattern resulting from a type-K hologram when (i) a reference plane wavefront is incident 
and (ii) when the information carrying wavefront is incident on the hologram.}
\label{fig_sp2}
\end{figure}
Mild external perturbation may cause movement of the beam incident on the type-K sensor. Such movement will lead to shift of 
the entire focal spot pattern captured by the camera. However any shift of the pattern may not be caused by external perturbation 
and some such shift may be due to the user encoded aberration mode present in the beam. In order to identify the externally 
caused beam movements, the transmission station can send a plane reference wavefront at periodic intervals. 
Such a reference wavefront will lead to a focal spot pattern, called reference focal spot pattern, such as the one seen in 
Fig. \ref{fig_sp2} (i). The type-K hologram is designed to generate a +1 order focal spot at the centre of the pattern which has the same wavefront as the reference wavefront. 
Thus the location of this focal spot indicates the centre of the focal spot pattern at a given instant and any change in the focal spot position between two successive reference 
focal spot patterns is a measure of the beam movement between the arrival of two reference wavefronts. In order to compensate for such beam movement, central intensities of the 
focal spots of the bias beams corresponding to the next information encoded wavefront (as the one seen in Fig. \ref{fig_sp2} (ii)) are to be extracted from locations shifted with 
respect to the previous information focal spot pattern by the same amount as the shift in the reference focal spot in the previous reference focal spot pattern. 
However in addition to the beam movement, the external perturbation may also incorporate an arbitrary distortion in the phase profile of the laser beam.
Since the Zernike modes form a complete set of basis functions, any arbitrary change in the phase profile caused by the external perturbation can be expressed as a linear combination of a subset of the Zernike modes. Let $\phi_R$ represents, phase distortions introduced by the perturbation, and hence the phase profile of the beam incident on the type-K sensor corresponding to the reference frame. We define $\phi_R=\sum\limits_q a_q Z_q$, where the summation may contain one or more sensor modes and Zernike modes other than the sensor modes. Therefore primary effect of perturbation will be the incorporation of additional strengths of the sensor modes which can be estimated from a measure of the type-K sensor output for each incident reference wavefront. Let $S_K^{ref}$ is the type-K sensor output obtained from the reference focal spot pattern such that $S_K^{ref}=S^s_K+S^i_K$, where $S^s_K$ is the contribution of the sensor modes present in $\phi_R$ and $S^i_K$ is the net contribution of modes other than the sensor modes present in $\phi_R$. If $S_K^{inf}$ is the sensor output corresponding to the next information focal spot pattern, then owing to the linearity of the type-K sensor we can write $S_K^{inf}=\phi^t_{RMS}+S_K^{ref}$. Here $\phi^t_{RMS}$ is (are) the root mean square value(s) of the sensor mode(s) transmitted from the transmission station carrying the user information. Therefore the correct value of the type-K sensor output for a given information frame is obtained as $S_K^{inf}-S_K^{ref}$.

$S_K^{inf}$ can be calculated after the beam movement has been compensated so that $S_K$ is corrected from both unwanted beam movements and the medium introduced sensor mode strengths.
The proposed compensation scheme however assumes that external perturbations do not change between the arrival of a reference wavefront and the next information carrying wavefront. For quickly changing perturbations the separation between the reference and the information carrying wavefront is to be reduced.
  
%

\end{document}